\documentclass[aps,twocolumn,nofootinbib,amsmath,amssymb,superscriptaddress]{revtex4}

\usepackage[utf8]{inputenc}
\usepackage{combelow} 
\usepackage{microtype}
\usepackage{graphicx}

\usepackage{natbib} 

\usepackage{color}
\usepackage{url}
\usepackage{hyperref}
\hypersetup{
  colorlinks   = true,    
  citecolor    = black,     
  linkcolor    = black,   
  urlcolor     = black    
}

\usepackage{multirow}
\usepackage{graphicx}
\usepackage{lscape}
\usepackage{booktabs}
\usepackage{textcomp}
\usepackage{array}
\usepackage{makecell}

\begin{document}

\title{HRTF Individualization: A Survey}

\author{\href{https://orcid.org/0000-0002-9823-335X}{\includegraphics[scale=0.06]{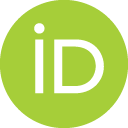}\hspace{1mm}Corentin \surname{Guezenoc}}}
\email{corentin.guezenoc@centralesupelec.fr}
\affiliation{3D Sound Labs SAS\\Rennes, France\vspace{1.2mm}}
\affiliation{FAST Research Team \\IETR (CNRS UMR 6164) \\CentraleSup\'elec \\Rennes, France}

\author{\href{https://orcid.org/0000-0001-7199-7563}{\includegraphics[scale=0.06]{orcid.png}\hspace{1mm}Renaud \surname{S\'eguier}}}
\email{renaud.seguier@centralesupelec.fr}
\affiliation{FAST Research Team \\IETR (CNRS UMR 6164) \\CentraleSup\'elec \\Rennes, France}

\begin{abstract}
	The individuality of head-related transfer functions (HRTFs) is a key issue for binaural synthesis. While, over the years, a lot of work has been accomplished to propose end-user-friendly solutions to HRTF personalization, it remains a challenge.
In this article, we establish a state-of-the-art of that work. We classify the various proposed methods, review their respective advantages and disadvantages and, above all, methodically check if and how the perceptual validity of the resulting HRTFs was assessed.


\end{abstract}

\maketitle

\section{Introduction}
Thanks to only two audio signals perceived at the eardrums, one is able to perceive the spatial characteristics of sound sources around him: distance, direction, spread... 
Among the auditory cues are the level, time-of-arrival and spectrum of the incoming sound. 
Typically, this sound/morphology interaction is mathematically described by the Head-Related Transfer Functions (HRTFs) \cite{moller_fundamentals_1992}.
These cues are greatly influenced by the interaction of sound with one's pinnae, head and torso 
and thus are specific to each individual. 

By reproducing these cues, a virtual auditory environment can be generated using regular headphones: by convolving a given sound sample with the right pair of HRTFs before presenting it to the listener, the sound sample is perceived at the desired location. This process is called binaural synthesis.
However, most binaural synthesis engines are currently non-individual, i.e. they use the same generic HRTF set for all users, which is known to cause discrepancies such as weak externalization, wrong perception of elevation and front-back inversions \cite{wenzel_localization_1993}. 
This is due to the fact that there is currently no easy way to provide individual HRTFs for the average customer.


Hence, an open key issue for binaural synthesis is: how to individualize HRTFs for the end-user? Furthermore, what is the perceptual performance of such an individualized HRTF set?
In this article, we go over the different families of approaches that address this problem, namely acoustic measurement, numerical simulation, indirect individualization based on morphological data and indirect individualization based on perceptual feedback. Furthermore, we systematically examine whether perceptual studies were conducted and what their results were and synthesize this information in Table \ref{tab-subj-results}.

\section{Acoustic Measurement}
The most obvious approach to HRTF individualization is acoustic measurement: one or several loudspeakers are positioned at each direction of interest around the subject and microphones placed at the entrance of his ear canals record the corresponding impulse responses. The measurement is usually performed in an anechoic or semi-anechoic environment (the HRTFs are, by definition, free-field transfer functions). Topics of interest include measurement setup, measurement time, subject-movement-related inaccuracies and, of course, perceptual performance.

\subsection{Measurement setup}
A typical state-of-the-art measurement setup \cite{rugeles_ospina_individualisation_2016, carpentier_measurement_2014, enzner_analysis_2008, mokhtari_toward_2008} features loudspeakers on one or several vertical arcs and a turntable on which the subject stands or sits, though a variety of measurement setups can be read of in the literature such as one or several loudspeakers moving around a still subject \cite{langendijk_fidelity_1999}.
This is the main shortcoming of the method: the equipment is expensive and scarcely transportable (and not at all in the case of anechoic or semi-anechoic measurements).
A more detailed presentation of measurement setups and their respective benefits and constraints can be found in Rugeles's PhD Thesis \citep[p. 46-49]{rugeles_ospina_individualisation_2016}.


\subsection{Measurement time}
Another major disadvantage of the method is the time needed to measure the HRTFs for thousands of directions. Indeed, between a few minutes and a couple of hours depending on the method, the subject is supposed to remain still for that duration, which is uncomfortable and difficult.
The historical approach, which consists in measuring the HRIRs one direction at a time, takes up to 1h45 on a modern setup such as Carpentier \emph{et al.}'s in 2014 \cite{carpentier_measurement_2014}. It is however often sped up, down to 20 mn according to Rugeles in 2016 \cite{rugeles_ospina_individualisation_2016}, using interleaved multiple sweep sines as proposed by Majdak \emph{et al.} in 2007 \cite{majdak_multiple_2007}.
A promising and rather trending approach is the one proposed by Enzner in 2008 \cite{enzner_analysis_2008}. Based on continuous azimuth-wise rotation and adaptive filtering, this new paradigm allowed the measurement time to be considerably reduced further: according to his work, it would only take 4 mn with that method to measure a whole HRIR set with a spatial resolution comparable to that of Rugeles's system \cite{rugeles_ospina_individualisation_2016}. 
\subsection{Directional imprecision due to subject movement}
Measurement time exacerbates another issue: as reported in 2010 by \cite{hirahara_head_2010} the subject cannot stay completely still all the way through the measurement session, which is a source of errors about the actual direction of the measured HRTFs (compared to the desired one). Nevertheless, recent studies \cite{majdak_3-d_2010, denk_controlling_2017} from 2010 and 2017 seem to have successfully limited the subject's movements by giving him a visual feedback. Denk \emph{et al.} \cite{denk_controlling_2017} reported their directional error to be imperceptible.
However, this directional imprecision at measurement might be an issue in several currently-used databases.

\subsection{Perceptual performance}
In spite of the aforementioned drawbacks of the method, for the last 30 years binaural synthesis with individual measured HRTFs has been extensively compared to real free-field sound sources in terms of localization accuracy. The consensus is that they are overall equivalent \cite{wightman_headphone_1989, moller_binaural_1996, langendijk_fidelity_1999, martin_free-field_2001, majdak_3-d_2010}, although a few defects \cite{wightman_headphone_1989} were reported and attributed either to the biasing presence of dynamic clues when comparing against real sources or to distortion in the measurements.
More details can be found in Bahu's PhD Thesis \citep[p. 27]{bahu_localisation_2016}.

\section{Numerical Simulation} \label{numerical_simulation}
Another approach to obtain an individual HRTF set is to simulate numerically the propagation of acoustic waves around the subject.
Its main advantages over HRTF measurement are mobility and user comfort. Indeed, only a 3D scan of the listener is needed for individualization which makes up for a much less tedious acquisition session than acoustic measurement. 
Moreover, once the 3D geometry is acquired, the simulation procedure is completely repeatable and free of measurement noise, and thus it holds a large potential to understanding the inter-individual variations in HRTFs.
Furthermore, a low-cost version can be made available to the end-user by using 2D-to-3D reconstruction techniques, by reducing the acquisition requirements to a set of consumer-grade smartphone pictures \cite{kaneko_deepearnet:_2016}.
Since the mid-2000s, the major computation techniques have been the Fast-Multipole-accelerated Boundary Element Method (FM-BEM)
\cite{gumerov_fast_2007, kreuzer_fast_2009, ghorbal_pinna_2017} 
for harmonic domain and the Finite Difference Time Domain (FDTD)
\cite{mokhtari_comparison_2007, prepelita_influence_2016} 
for time domain, though other methods such as the Finite Element Method (FEM) \cite{huttunen_simulation_2007} and the more exotic raytracing \cite{rober_hrtf_2006} and Differential Pressure Synthesis (DPS) \cite{tao_differential_2003} have been used since the late 1990s, 2006 and 2003, respectively.
We take a particular interest here into the matters of the accuracy of the 3D geometry used for simulation, the computing time and the perceptual relevance of the calculated HRTFs.
 

\subsection{3D Geometry Accuracy}
A major topic of interest for HRTF calculation is the accuracy of the 3D geometry passed into simulation.

Therefore, geometry acquisition is a key issue. On this, there seems to be a consensus on the fact that the ear needs more accuracy than the rest of the bust. Typically, a precise scan of the ear is stitched onto a rougher scan of the head and/or torso by an operator, which takes up to dozens of minutes of manual labour. A wide variety of scanning solutions can be read of in work on HRTF calculation: MRI, CT scan, structured light and infrared for instance. Scanning of the pinna have sometimes been performed on a mold. 
However, the literature would merit more studies that evaluate and compare the various scanning methods and their impact on the resulting HRTFs. 

In contrast, the matter of geometry re-meshing has been well-studied.
Indeed, prior to BEM simulation, the surfacic mesh of the subject must be re-arranged so it is regular enough and so the edge lengths are small enough in regard to the simulation's wavelength. As computing time increases considerably with the number of mesh elements, the re-meshing resolution is a trade-off between numerical accuracy and computing time.
Although the use of the six-to-ten-elements-per-wavelength empirical rule 
has been wide-spread, the Acoustics Research Institute has recently well contributed to the subject. 
Indeed, by implementing and studying the effect of various re-meshing methods on the resulting HRTFs objectively and subjectively, they not only determined the optimal uniform re-meshing resolution in 2015 \cite{ziegelwanger_numerical_2015} but also proposed a progressive re-meshing algorithm that allowed the simulation time to be cut down by a factor 10 while maintaining the same HRTF accuracy in 2016 \cite{ziegelwanger_priori_2016}.
%
%
Similar work has been carried out in the case of FDTD simulation through studying the impact of the voxelization of a subject's volumic geometry on the resulting HRTFs \cite{prepelita_influence_2016}.

\subsection{Computing time}
Computing time used to be the main drawback of HRTF calculation: HRTFs could not be computed on the whole audible frequency range up until 2007 \cite{huttunen_simulation_2007, mokhtari_comparison_2007}. However, it has been reduced to a few hours' time thanks to the constant increase in available computing power, to the democratization of distributed computing on clusters over the last decade and to the introduction of FM-BEM in 2007 \cite{gumerov_fast_2007}.

\subsection{Perceptual Performance}
Various objective comparisons with acoustic measurements reported computed HRTF sets to be overall similar to acoustic measurements \cite{gumerov_fast_2007, kreuzer_fast_2009, ziegelwanger_calculation_2013}, 
although one of them \cite{kreuzer_fast_2009} reported some alterations of spectral features known to be clues for elevation perception.
On a subjective level, among the studies where individual HRTF sets were simulated for human subjects on the whole audible range (i.e. up to at least 16 kHz), two provided perceptual evaluations \cite{mokhtari_toward_2008, ziegelwanger_numerical_2015}. Mokhtari \emph{et al.} in 2008 \cite{mokhtari_toward_2008} and Ziegelwanger \emph{et al.} in 2015 \cite{ziegelwanger_numerical_2015} performed localization tests with measured HRTFs as reference that showed good results, however these studies were carried out on very few subjects: 2 and 3 respectively.

\section{Indirect Individualization based on Anthropometric Data}
\label{indiv-morpho}
Though more convenient than acoustic measurement, HRTF calculation still requires specialized equipment and non-negligible mesh processing and computing time. 
Hence, based on the fact that HRTF sets rely heavily on morphology, 
many studies have explored the idea of a low-cost HRTF individualization methodology based on anthropometric measurements. We distinguish three sub-categories: adaptation, selection and regression.


\subsection{Adaptation} \label{adaptation_morpho} 
One way to do it is to take a non-individual set and to adapt it, i.e. to alter it in order to make more suitable for the subject at hand.
Based on the idea that the most prominent morphological difference between two individuals is size, Middlebrooks and colleagues \cite{middlebrooks_individual_1999} proposed in 1999 to adapt a generic HRTF set thanks to a frequency scaling. In 2000 \cite{middlebrooks_psychophysical_2000}, they reported that the scaling factor could be estimated from a combination of head and pinnae measurements through linear regression. In both cases, perceptual evaluations performed on 9 to 11 subjects reported localization performance to be improved compared to no individualization but to be worse than with own measured HRTF set.
Later on in 2005 and 2008, other researchers \cite{maki_reducing_2005, guillon_head-related_2008} combined frequency scaling with a rotation in space of the HRTF set, which translates to a head tilt, in order to further improve the adaptation's results. However, neither of these studies included any perceptual study. In particular, it was impossible to Maki \emph{et al.} \cite{maki_reducing_2005} to do so as the HRTFs they studied were those of gerbils.


\subsection{Selection}
Complementary to adaptation, one can select a HRTF set from anthropometric measurements in a database that contains both kind of data. 
For instance, using the CIPIC database \cite{algazi_cipic_2001}, Zotkin \cite{zotkin_customizable_2002} implemented in 2002 a coarse nearest neighbors approach that used only 7 morphological parameters measured on a picture of the pinna, and showed some improvement in terms of localization performance compared to no individualization (average gain of $15 \%$ in elevation score).
More recently, in 2017, Yao \cite{yao_head-related_2017} proposed a more exotic method to select a HRTF set among a database, using a neural network trained to predict a perceptual score (from 1 to 5) from anthropometric measurements. However, it is difficult to conclude on the results of their perceptual study in comparison with others, as it only used their own perceptual score as indicator.



\subsection{Regression}
Going further, another approach to devising low-cost HRTF individualization based on morphology is the estimation of a HRTF set from anthropometric measurements of the listener.
To this end, multiple linear regression has been widely used. 
Among such work, the HRTF sets have often, since the early 2000s, been compressed using statistical modeling such as Principal Component Analysis (PCA) \cite{jin_enabling_2000, hu_head_2006}
and Independent Component Analysis (ICA) \cite{huang_hrir_2009}.
Some, as Bilinski \emph{et al.} in 2014 \cite{bilinski_hrtf_2014}, have chosen to rather predict a HRTF set by linear combination of HRTF sets using the coefficients of a model of anthropometric parameters.
Suprisingly, among the studies reviewed for this article, only that of Hu \emph{et al.}\cite{hu_head_2006} featured a perceptual evaluation and, while the results were encouraging, they did not put elevation perception to the test.
Since the late 2000s, nonlinear regression models have been used too that have typically relied on neural networks coupled to various data compression techniques including PCA, 
\cite{hu_hrtf_2008} 
High-Order SVD \cite{li_hrtf_2013} and Isomap \cite{grijalva_anthropometric-based_2014}. However, none of these studies carried out any perceptual evaluation of the estimated HRTF sets.

%

\section{Indirect Individualization based on Perceptual Feedback}
If methods for indirect individualization based on morphological data are practical for the end-user and provide individualization, they can be subject to morphological measurement errors. Indeed, the morphological data acquisition is done by the user: measurements as well as pictures can be made wrong. 
As the subjective perception of spatialization is the ultimate goal, an alternative is to propose a low-cost individualization method that is based on the listener's feedback. 
Quite similarly to section \ref{indiv-morpho}, we distinguish two categories: selection and adaptation.


\subsection{Selection}
A natural strategy that has been well-explored in the literature since the late 1990s is to help the listener select the best non-individual HRTF set among a database \cite{seeber_subjective_2003, katz_perceptually_2012}. 
All studies reviewed for this article evaluated the selected HRTF set perceptually with results indicating that the selected set was better than a non-individual one but worse than a subject's own set. However, it should be noted that Seeber \emph{et al.} \cite{seeber_subjective_2003}
did not put elevation perception to the test in their study. Reported tuning times ranged from 15 min \cite{seeber_subjective_2003} 
 to more than 35 min \cite{katz_perceptually_2012}.
Conjointly, in order to improve the relevance and duration of the tuning procedure, it has been proposed to cluster \textit{a priori} the database based on either objective \cite{xie_typical_2015} or perceptual \cite{katz_perceptually_2012} criteria.


\subsection{Adaptation}
A non-individual HRTF set, sometimes elected through a previous selection procedure, can be adapted based on perceptual feedback from the listener. We distinguish three ways to adapt a HRTF set: frequency scaling, filter-design-based tuning and statistical-model-based tuning.

%

\subsubsection{Frequency scaling}
As mentioned in \ref{adaptation_morpho}, Middlebrooks \emph{et al.} explored in 1999 \cite{middlebrooks_individual_1999} the idea of adapting a generic HRTF set through frequency scaling and reported in its companion study \cite{middlebrooks_virtual_1999} an improvement in localization performance compared to no scaling. In their 2000 study \cite{middlebrooks_psychophysical_2000}, they reported that the scaling factor could be tuned by the listener trough a 20-min tuning session with similar localization performance than previous methods for obtaining the scaling factor (minimization of a spectrum-based metric and anthropometric measurements). 
This tuning method has the advantage of offering one single tuning lever for the whole HRTF set and to bring some perceptual improvement.

%

\subsubsection{Filter-design-based tuning}
Some work \cite{tan_user-defined_1998, runkle_active_2000} proposed in 1998 and 2000, respectively, to rely on the tuning of filters to adapt a given HRTF set. 
We have distingushed two directions. First, direction dependance was not handled \cite{tan_user-defined_1998}, which meant the adaptation was rather rough as it is basically an equalization of the whole HRTF set. Second, the listener-driven filter-design had to be done for each direction separately \cite{runkle_active_2000} and thus the number of parameters to tune for a whole set was too high to expect a tuning procedure in a reasonable amount of time.
Indeed, Runkle \emph{et al.} \cite{runkle_active_2000} did not present any perceptual evaluation of their solution while Tan and Gan \cite{tan_user-defined_1998} presented some encouraging perceptual results but did not evaluate other criteria that the ones used for tuning i.e. front-back reversal and sense of elevation.

%

\subsubsection{Statistical-model-based tuning}
Alternatively, a lot of work have proposed to rely on a statistical model, with in mind the goal of reducing the number of tuning parameters while still being able to cover most of the database's HRTF space.

The main statistical modeling method used in the literature is Principal Component Analysis (PCA) for its ease to interpret as well as for its low implementation and computing complexity. 
Most \cite{shin_enhanced_2008, hwang_modeling_2008-1, fink_individualization_2015}, in 2008, 2008 and 2015 respectively,  proposed a procedure that allowed the tuning of a HRTF in one direction at a time. The number of parameters were reduced to 3 to 5 principal components (PC) weights per direction, making it possible for the listener to tune each direction in a reasonable amount of time. These studies all reported a localization performance improvement over non-individual HRTFs, although the number of subjects was rather small (3 and 4 respectively) for \cite{shin_enhanced_2008} and \cite{hwang_modeling_2008-1} and elevation perception was not evaluated in \cite{fink_individualization_2015}. However, these tuning procedures had to be performed direction by direction and thus did not allow to tune a whole HRTF set in a reasonable amount of time (only 9 to 10 directions were tuned).
Hölzl, in his 2014 Master Thesis \cite{holzl_global_2014}, proposed a solution to that flaw by applying Spherical Harmonics (SH) to the direction-dependent PC weights. However, no subjective evaluation of this method was proposed, and even though the overall problem dimension was reduced to 5 PC weights x 9 SH coefficients = 45, it is still a high number of parameters to tune. Moreover, the combination of spherical harmonics coefficients and principal component weights are rather counter-intuitive and hard to comprehend for the end-user.

In 2017, Yamamoto and Igarashi \cite{yamamoto_fully_2017} proposed a state-of-the-art method that relied on the modeling of HRTF sets thanks to a variational autoencoder neural network. The tuning procedure consisted in a gradient descent optimization of the network's weights where the cost was determined at every iteration by the user's notation of two HRTF sets presented to him by the algorithm. 
They conducted a preference test in which the participants graded HRTF sets pair by pair in a double-blind manner. The baseline condition was a best fit non-individual HRTF set elected among the database in a previous preference test procedure. The outcome was a significant improvement over an optimal non-individual HRTF set for 18 participants out of 20, although the nonstandard nature of the perceptual testing methodology makes it hard to compare those results with other studies'.

\section{Discussion}

\newcolumntype{L}[1]{>{\raggedright\let\newline\\\arraybackslash\hspace{0pt}}p{#1}}
\newcolumntype{C}[1]{>{\centering\let\newline\\\arraybackslash\hspace{0pt}}p{#1}}
\newcolumntype{R}[1]{>{\raggedleft\let\newline\\\arraybackslash\hspace{0pt}}p{#1}}
\newlength{\firstcolwidth} \setlength{\firstcolwidth}{.3\textwidth}
\newlength{\secondcolwidth} \setlength{\secondcolwidth}{.115\textwidth}
\newlength{\fifthcolwidth} \setlength{\fifthcolwidth}{.08\textwidth} 
\newlength{\sixthcolwidth} \setlength{\sixthcolwidth}{.23\textwidth}

\begin{table*} 
\centering

\begin{tabular}{@{}
				L{\firstcolwidth}
				L{\secondcolwidth}
				c 
				c 
				C{\fifthcolwidth}
				L{\sixthcolwidth}
				@{}}
\toprule 
	& Eval. type
	& Baseline
	& $N_{subj}$ 
	& $\tau_{perc}$ (\%)
	& Results
	\\ \midrule
\textbf{Acoustic mesurement}  
\cite{wightman_headphone_1989, moller_binaural_1996, langendijk_fidelity_1999, martin_free-field_2001, majdak_3-d_2010} 
	& Localization 
	& RS
	& 3-10 
	& \newline 63 
	& \newline Good 
	\\ \addlinespace 
	& Preference 
	& RS
    & 6 
	& 
	& 
	\\ \addlinespace \midrule
\textbf{Numerical simulation} 
\cite{mokhtari_toward_2008, ziegelwanger_numerical_2015}
	& Localization  
	& IAC
	& 3
	& 25 
	& Promising but would merit more studies \& subjects 
	\\ \addlinespace \midrule
\textbf{Indirect individualization from anthropometric data}
	&
	&
	&
	&
	&
	\\ \addlinespace
Selection, frequency-scaling-based adaptation
\cite{middlebrooks_psychophysical_2000, yao_head-related_2017}
	& Localization 
	& NIAC
	& 6-11 
	& 67 
	& Better than non-individual 
	\\ \addlinespace 
Statistical-model-based regression 
\cite{hu_head_2006} 
	& Localization, no elevation 
	& NIAC
	& 5
	& 10 
	& Poor: few studies and no elevation testing
	\\ \addlinespace \midrule
\textbf{Indirect individualization from perceptual feedback}
	&
	&
	&
	&
	&
	\\ \addlinespace
\multirow{2}{\firstcolwidth}
	{Selection, frequency-scaling-based adaptation
	 \cite{middlebrooks_psychophysical_2000, seeber_subjective_2003, katz_perceptually_2012}
	}
	& Localization \newline 
	& NIAC
	& 7-11 
	& \newline 100 
	& \multirow{2}{\sixthcolwidth} 
		{\newline Better than non-individual}
 	\\ \addlinespace 
	& Preference \newline 
	& NIAC 
    & 45 
	& 
	&  
	\\ \addlinespace 
\multirow{2}{\firstcolwidth}
	{Filter-design-based adaptation, statistical-model-based adaptation
	\cite{shin_enhanced_2008, hwang_modeling_2008-1, fink_individualization_2015, yamamoto_fully_2017}
	}
 	& Localization \newline 
 	& IAC, NIAC 
 	& 3-6 
 	& \newline 80 
 	& \multirow{2}{\sixthcolwidth}
 		{Promising but would merit more standard studies \& more subjects} 
 	\\ 
	& Preference 
	& BFAC
    & 20
	& 
	& 
 	\\ \bottomrule 

\end{tabular}

\caption{Overview of perceptual evaluations for the major HRTF individualization approaches. 
\\The columns describe the following features, from left to right: type of evaluation (Eval. type), condition(s) used as ground truth (Baseline), number of participants ($N_{subj}$), proportion of studies that carried out a perceptual evaluation ($\tau_{perc}$) and results of the perceptual studies. 
\\Acronyms RS, IAC, NIAC and BFAC stand respectively for Real sound Sources, stimuli binauralized using Individual Acoustic HRTFs, stimuli binauralized using Non-Individual Acoustic HRTFs and stimuli binauralized using a Best Fit non-individual Acoustic HRTF set elected among the database in a previous preference test procedure.
}
\label{tab-subj-results}

\end{table*}

As of today, acoustic measurement remains the reference method to acquire individual HRTFs thanks to significant perceptual assessment against real sound sources \cite{wightman_headphone_1989, moller_binaural_1996, majdak_3-d_2010}, as summarized in Table \ref{tab-subj-results}. As such, it has been used as ground truth by all other families of HRTF individualization methods. Nevertheless, in spite of recent major advances in terms of acquisition time, it is impractical for consumer-grade applications because of the cost and difficulty to transport the measurement equipment.

On the other hand, in spite of the professional-grade scanning equipment and few processing hours needed 
, numerical simulation allows the data acquisition step to be mobile and more comfortable for the user. Furthermore, the scanning equipment may be reduced to a simple smartphone for consumer-grade applications by relying on 2D-to-3D reconstruction technologies\cite{kaneko_deepearnet:_2016}.
In addition, simulation is a powerful tool for investigating and understanding the link between morphology and HRTFs. 
Major technical limitations such as computing time, 3D geometry acquisition and re-meshing have mostly been overcome. However, although objective \cite{kreuzer_fast_2009, gumerov_fast_2007, ziegelwanger_calculation_2013} and subjective \cite{mokhtari_toward_2008, ziegelwanger_numerical_2015} evaluations showed rather promising results, perceptual studies that compared calculated HRTFs with measured ones were surprisingly rare and featured only 2 to 3 subjects (cf Table \ref{tab-subj-results}. In addition, some objective observations underlined the possibility of perceptual defects in the produced HRTFs.
Hence, despite a lot of work on HRTF simulation for thirty years, and in particular since the first full-band calculations ten years ago, computed HRTFs would merit wider-ranged perceptual studies, both in number of studies and of participants.
Possible causes for simulation-related problems include an inaccurate geometry acquisition (depending on the scanning process) and/or a wrong modeling of the acoustics problem.

With in mind the goal of developing solutions that are more user-friendly, the idea of individualizing HRTFs from simpler morphological data has been widely explored in the literature. This has the advantage of relying on little equipment and on an easy data acquisition process, usually a smartphone and the shooting of one or a few pictures. 
However, as reported in Table \ref{tab-subj-results}, the perceptual results are mixed.
On one side, the simple methods, namely selection and adaptation by frequency scaling and/or set rotation, have demonstrated some perceptual improvement compared to no individualization, thanks to studies that featured 6 to 11 participants \cite{middlebrooks_psychophysical_2000, yao_head-related_2017}.
On the other side, we cannot conclude on the quality of the HRTFs produced by more complex methods, such as linear and nonlinear regression between anthropometric measurements and HRTF sets. Indeed, among the last category we found a rare single perceptual study \cite{hu_head_2006} and that one did not try elevation perception. In other words, there is a lack of perceptual results for statistics-based methods, which may well indicate that the databases are not large enough: all the studies reviewed here used similarly-sized databases of 43 to 50 subjects. Thus, a key to their improvement may well reside in larger databases. However, to the best of our knowledge the matter of their ideal size remains an open one.
More generally for the anthropometrics-based approach, errors may also come from the fact that the measurement step is handed over to the end-user and from the unclear relevance of the choice of the anthropometric parameters to predict HRTFs. 
  

Alternatively, researchers have investigated the possibility of individualizing a HRTF set based on the listener's subjective feedback. This approach has the double advantage of including the listener and his perceptions in the individualization process while avoiding errors related to data acquisition. Accordingly, the vast majority of such studies provide subjective evaluations (cf Table \ref{tab-subj-results}).
On one hand, the simple techniques, which include selection and adaptation by frequency-scaling, have shown perceptual improvement over no individualization in studies that gathered 7 to 11 listeners \cite{middlebrooks_psychophysical_2000, seeber_subjective_2003}. On the other hand, the more complex methods i.e. the statistical-model-based ones, have been well used in order to reduce the number of tuning parameters in the most relevant manner. 
To this end, PCA models have been used in majority \cite{shin_enhanced_2008, hwang_modeling_2008-1, fink_individualization_2015}. While the models that were used needed to be tuned direction by direction and thus the tuning of a whole HRTF set was impractical, they have shown encouraging results to their localization tests, though some \cite{shin_enhanced_2008, hwang_modeling_2008-1} featured only 3 to 4 subjects and the other \cite{fink_individualization_2015} only included azimuthal directions.
As for Yamamoto and Igarashi \cite{yamamoto_fully_2017}, the result of their 20-listener preference test was altogether promising, but it would merit a more standard subjective evaluation to be able to compare it to other studies.
For further advances, statistical-model-based approaches, as in the case of anhtropometry-based indirect methods, may very well benefit from larger databases. Indeed, it would then be particularly interesting to attempt PCA modeling of whole HRTF sets and to use its weights as tuning parameters. Yamamoto and Igarashi's \cite{yamamoto_fully_2017} method seems promising as well but would benefit from a more conventional perceptual evaluation methodology such as localization testing.




\section{Conclusion}

%
%

In this paper we established a state-of-the-art of what has been done so far to tackle the problem of HRTF individualization for the end-user. We distinguished four families of methods, namely acoustic measurement, numerical simulation, indirect individualization from morphology and indirect individualization from perceptual feedback. We summarized their specific advantages and disadvantages and took stock of the current advances while identifying some leads for improvement. 
In particular, we took a special interest into the existence and outcome of related perceptual studies. Overall, significant perceptual results are rather scarce, though not for all approaches (cf Table  \ref{tab-subj-results}), which tends to indicate that a lot of work remains to be done to reach an efficient end-user-friendly solution to HRTF individualization.


%

\bibliography{2018-08-03_zotero}

\end{document}